\documentstyle[preprint,aps,epsf]{revtex}
\newif\iftightenlines\tightenlinesfalse
\tightenlines\tightenlinestrue

\def\to{\rightarrow}

\def\te{\tilde e}

\def\tt{\tilde t}
\def\tu{\tilde u}

\def\tb{\tilde b}

\def\td{\tilde d}

\def\tst{\tilde t}
\def\ttau{\tilde \tau}

\def\tg{\tilde g}
\def\tnu{\tilde\nu}
\def\tell{\tilde\ell}

\def\tw{\widetilde W}
\def\tz{\widetilde Z}
\begin{document}
\draft
\preprint{\vbox{\baselineskip=14pt%
   \rightline{FSU-HEP-000102}
   \rightline{UH-511-950-00}
}}
\title{CALCULABLE SPARTICLE MASSES WITH\\
RADIATIVELY DRIVEN INVERTED MASS HIERARCHY}
%
\author{Howard Baer$^1$, Pedro Mercadante$^1$ and Xerxes Tata$^2$}
\address{
$^1$Department of Physics,
Florida State University,
Tallahassee, FL 32306 USA
}
\address{
$^2$Department of Physics and Astronomy,
University of Hawaii,
Honolulu, HI 96822, USA
}
\date{\today}
\maketitle
\begin{abstract}

Supersymmetric models with an inverted mass hierarchy (IMH: multi-TeV
first and second generation matter scalars, and sub-TeV third 
generation scalars) can ameliorate problems arising from 
flavor changing neutral currents, $CP$ violating phases and 
electric dipole moments, while at the same time satisfying 
conditions on naturalness. It has recently been shown that such an
IMH can be generated radiatively, making use of 
infra-red fixed point properties of renormalization group equations
given Yukawa coupling unification and suitable $GUT$ scale boundary 
conditions on soft SUSY breaking masses. In these models, explicit spectra
cannot be obtained due to problems implementing radiative 
electroweak symmetry breaking (REWSB). We show that use of $SO(10)$ $D$-term
contributions to scalar masses can allow REWSB to occur, while maintaining
much of the desired IMH. A somewhat improved IMH
is obtained if splittings are applied only to Higgs scalar masses.
\end{abstract}

\medskip



Weak scale supersymmetry is 
especially appealing because destabilizing quadratic divergences that
are present in the Standard Model cancel upon the 
introduction of supersymmetry\cite{susy}. Constraints from naturalness 
generally restrict superpartner masses to be below $\sim 1$ TeV, so that
these new states of matter ought to be accessible to present or near future
collider experiments\cite{nat}. In general, 
however, there also exist constraints
on superpartner masses from flavor changing neutral current
processes\cite{fcnc} ({\it e.g.} in the $K-\bar{K}$ system), 
electric dipole moments 
of the electron and neutron\cite{edm}, 
$\mu \to e\gamma$ decays\cite{meg}, proton decay\cite{pd}
and Big Bang nucleosynthesis\cite{bbn}, 
that all favor superpartner masses in the
multi-TeV range, unless specific assumptions such as universality\cite{dg}
of scalar masses, or alignment\cite{align} 
between fermionic and bosonic mixing matrices,
are made. These considerations have provided a strong motivation
to construct new and interesting mechanisms for the communication of 
supersymmetry breaking\cite{gmsb,amsb}.

It is important to notice that 
naturalness arguments most directly apply
to third generation superpartners, owing to their large 
Yukawa couplings. In contrast, the constraints from flavor physics
mentioned above
apply (mainly) to scalar masses of just the first two generations.
This observation has motivated the construction of a variety of models,
collectively known as inverted mass hierarchy (IMH) models\cite{imh}, 
where the first and second generation squarks and sleptons have multi-TeV
masses, while third generation scalars have sub-TeV masses.
If the IMH already applies at or near the scale of grand unification,
then it has been shown\cite{mur} that two-loop contributions to 
renormalization group (RG) running cause tachyonic third generation 
squark masses to occur, unless these masses are beyond $\sim 1$ TeV,
which again pushes the model towards the ``unnatural''.

A resolution of this GUT scale dilemma has been presented recently in a series
of papers{\cite{rimh,bfpz} where the theme 
is that the IMH can be generated radiatively
by starting with {\it all} scalar masses at the multi-TeV level
at or near the GUT scale. 
It is pointed out that the infrared fixed point behaviour of the RG
equations, together with a simple choice of boundary conditions, results
in sub-TeV masses in the gaugino/Higgsino and third generation scalar
sectors, while the first two generations of scalars are left in the
multi-TeV range.
An essential ingredient in the analysis is the presence of a 
singlet neutrino superfield $\hat{N}^c$ in addition to the 
usual superfields of the MSSM. The right-handed neutrino is 
expected to decouple at intermediate scales $Q=10^{11}-10^{13}$ GeV,
leading to eV scale masses for the tau neutrino 
via the see-saw mechanism as is suggested by recent atmospheric neutrino
data~\cite{superK}

The most refined set of GUT scale boundary conditions\cite{bfpz} stipulate that
\begin{equation}
4m_Q^2=4m_U^2=4m_D^2=4m_L^2=4m_E^2=4m_N^2=2m_{H_u}^2=2m_{H_d}^2=A_0^2,
\label{eq1}
\end{equation}
consistent with minimal SUSY $SO(10)$ unification.
With boundary values for these parameters in the multi-TeV range,
large suppression factors are
generated for third generation and Higgs scalars, while other matter
scalars remain heavy.
The authors note that the radiatively driven IMH model has
a problem with generating an appropriate radiative electroweak
symmetry breaking (REWSB), which is common to all models at such
high values of $\tan\beta \sim 50$, where Yukawa couplings 
most nearly unify.
In some of the examples presented in Ref. \cite{rimh} and \cite{bfpz}, 
the squared Higgs masses
never reach the negative values required by REWSB, while in other examples,
various matter scalar squared masses are driven negative.
Without an explicit mechanism for electroweak symmetry breaking, it is
not possible to obtain mass spectra and couplings within this new and
interesting picture. 

In a recent paper\cite{bdft}, it has been shown that explicit mass spectra
can be calculated in Yukawa unified $SO(10)$ models consistent with
REWSB if $D$-term contributions to scalar masses are
included at $Q=M_{GUT}$. The $D$-term contributions are expected
whenever spontaneous symmetry breaking reduces the rank of the gauge
group as {\it e.g.} when $SO(10)$ breaks to
$SU(3)\times SU(2)\times U(1)$. Solutions with Yukawa coupling 
unification good to 5\% were found
only for positive $D$ terms, which caused a split $m_{H_d}>m_{H_u}$
at the GUT scale, and only for negative values of the superpotential $\mu$
parameter. 

In this paper, we examine the analogous solution to the problem of REWSB
in the IMH model. We note that the analytic derivation of the GUT scale
boundary condition discussed above used~\cite{bfpz} one-loop RGEs applied
only to soft SUSY breaking mass parameters that acquire multi-TeV scale
masses. Terms in the RGEs proportional to sub-TeV quantities such as
gaugino mass parameters were argued to be small, and dropped. In a
realistic calculation, weak scale parameters and two-loop contributions
to RGEs (if used) will modify the IMH solution somewhat. The expectation is
that the desired qualitative features of an IMH will survive these additional
perturbations. Our hope here is that there is some range of parameters
for which a limited
non-universality in scalar masses (originating in the $D$-terms) will
allow REWSB to occur, while not spoiling too much of the expected
IMH. 

In our initial set of calculations, we assume that an $SO(10)$
SUSY GUT breaks to the MSSM plus a right-handed neutrino (MSSM+RHN)
at a scale $Q=M_{GUT}$.
At this scale, the scalar squared masses are given by
\begin{eqnarray*}
m_Q^2&=&m_E^2=m_U^2=m_{16}^2+M_D^2 \\
m_D^2&=&m_L^2=m_{16}^2-3M_D^2 \\
m_N^2&=&m_{16}^2+5M_D^2 \\
m_{H_{u,d}}^2&=&m_{10}^2\mp 2M_D^2 , \\
\end{eqnarray*}
where $M_D^2$ parametrizes the magnitude of the $D$-terms, and can,
owing to our ignorance of the gauge symmetry breaking mechanism, be
taken as a free parameter, with either positive or negative values.
Here, $m_{16}$ denotes the common mass of the 16-component spinor
representation of $SO(10)$ to which the matter scalars including the
right sneutrino belong, while $m_{10}$ denotes the mass of the
10-component representation that contains the two Higgs doublets of the
MSSM. The model is completely specified by the parameter set,
\begin{eqnarray*}
m_{16},\ M_D^2,\ m_{1/2},\ M_N,\ \tan\beta ,\ sign(\mu ),
\end{eqnarray*}
where $m_{10}$ and $A_0$ are determined in terms of $m_{16}$ by the
boundary condition above.

To calculate the superparticle and Higgs boson mass spectra, we adopt 
the bottom-up approach inherent in ISASUGRA, which is a part of the 
ISAJET program\cite{isajet}.
Our procedure is as follows. We generate random samples of model
parameters
\begin{eqnarray*}
1000&<&m_{16}<10000\ {\rm GeV},\\
0&<&m_{1/2}<1000\ {\rm GeV},\\
0&<&M_D^2<m_{16}^2/3,\\
10&<&\tan\beta <55, \\
\mu &>&0\ {\rm or}\ \mu <0 ,
\end{eqnarray*}
while allowing $M_N$ to float between $5\times 10^{12}$ and 
$5\times 10^{13}$ GeV.

Starting
with the three gauge couplings and $t$, $b$ and $\tau$ Yukawa couplings
of the MSSM at scale $Q=M_Z$ (or $m_t$),
ISASUGRA evolves the various couplings up
in energy until the scale where $g_1=g_2$,
which is identified as $M_{GUT}$, is reached. The $GUT$
scale boundary conditions are imposed, and the full set of RGEs
for gauge couplings, Yukawa couplings and relevant scalar masses are evolved
down to $Q\sim M_{weak}$, where the renormalization group improved
one-loop effective potential is minimized at an optimized scale choice
$Q=\sqrt{m_{\tst_L}m_{\tst_R}}$ and radiative electroweak symmetry breaking
is imposed. Using the new spectrum, the full set of SSB masses and couplings
are evolved back up
to $M_{GUT}$ including weak scale sparticle threshold corrections to
gauge and Yukawa couplings. 
The process is repeated iteratively until a stable solution
is obtained. We use one loop RGEs for the soft SUSY breaking
parameters, but two-loop equations for the gauge and Yukawa couplings.

Our first results are shown in Fig. \ref{imh1}. Here, we plot
solutions to the superparticle mass spectrum consistent with REWSB, for
$\mu <0$ and $A_0 <0$.
The quantity $S$ is the ``crunch'' factor defined as, 
\begin{eqnarray*}
S   =
 {{3(m_{u_L}^2+m_{d_L}^2+m_{u_R}^2+m_{d_R}^2)+m_{\te_L}^2+m_{\te_R}^2+
m_{\tnu_e}^2}
 \over 
{3(m_{\tt_1}^2+m_{\tb_1}^2+m_{\tt_2}^2+m_{\tb_2}^2)+m_{\ttau_1}^2+m_{\ttau_2}^2+
m_{\tnu_{\tau}}^2}}.
\end{eqnarray*}
Notice that this differs slightly from the corresponding definition in
Ref.\cite{bfpz} since we are able to use mass eigenvalues
in the definition.
In frame {\it a}),
we show $S$ versus the ratio $M_D/m_{16}$. We see that all solutions
have {\it some} suppression of third generation masses, due to the large
Yukawa couplings of the third generation. However, for $M_D/m_{16}\sim
0.2$, the value of $S$ can reach values as high as 6-7. Indeed we see
that most solutions have a significantly smaller value of $S$. This is
in part due to our non-requirement of Yukawa coupling unification,
which was assumed in the derivation of (\ref{eq1}).
In frame {\it b}), we show $S$ versus $\tan\beta$. Here, it is easy to
see that a maximum IMH develops for very large values of $\tan\beta$
where Yukawa unification can occur.  The remaining frames show $S$
versus a ratio indicating the degree of Yukawa coupling unification
$R_{tb}=|(f_t-f_b)/f_t|$, where $f_t$, $f_b$ and $f_\tau$ are the third
generation Yukawa couplings evaluated at $Q=M_{GUT}$.  $R_{\tau b}$ is
similarly defined. In frames {\it c}) and {\it d}), we see that the
maximum IMH is indeed obtained typically for the smaller values of $R$,
where Yukawa couplings are most nearly unified.  Similar results and
suppression factors are obtained for $\mu >0$ solutions, although in
this case Yukawa coupling do not unify as well as for $\mu < 0$.

Two examples of specific spectra with considerable $S$ factors and full
$SO(10)$ $D$-terms are shown as case 1 and case 2 in Table 1. Case 1 has
$\mu <0$ and case 2 has $\mu >0$.  In case 1, first generation scalar
masses are $\sim 1500$ GeV, while the lightest third generation squarks
are $m_{\tb_1}=310.9$ GeV and $m_{\tst_1}=364.7$ GeV. The $\tb_1$ is the
lightest third generation scalar, and is just beyond the region
accessible to searches at
the Fermilab Tevatron\cite{bmt}.  The $\tw_1\to \tz_1 W$ at $\sim
100$\%, while $\tz_2\to \tz_1 Z^0$ or $\tz_1 h$ dominantly. 
In case 2, first
generation scalars again have $m\sim 1500$ GeV, but in this case the top
squark is the lightest third generation scalar ($m_{\tst_1}=219.3$ GeV).
The $\tw_1$, with
$m_{\tw_1}=124.1$ GeV, should be accessible to Fermilab Tevatron
searches\cite{trilep,invert}, since the $\tw_1$ and $\tz_2$ decay via
three-body modes which are dominated by $W$ and $Z$ exchange diagrams,
so leptonic branching fractions are not suppressed.

Applying the $SO(10)$ $D$-terms to scalar masses upsets the precise form
of the boundary condition of Eq. \ref{eq1}, and the values of $S$ we
obtain fall short of what has been obtained in Ref. \cite{bfpz}.  
We also examined
whether higher $S$ values can be achieved by applying splitting 
only to the Higgs
squared masses (since it is this splitting which allows for REWSB),
leaving the matter scalars degenerate at $M_{GUT}$. While inconsistent
with the $SO(10)$ framework, it adheres more closely to the boundary
conditions in Eq. \ref{eq1}.  We continue to parameterize the mass
splitting in terms of the parameter $M_D$. These results may be relevant
for scenarios with smaller GUT groups, but with a singlet neutrino.  Our
results with only splitting in the Higgs masses are presented in
Fig. \ref{imh2}, again for $\mu <0$ and $A_0 <0$.  We see in frame {\it
a}) that in this case, somewhat larger $S$ values up to $8-9$ can be
obtained, but typically for values of $M_D/m_{16}\sim 0.4-0.6$. In frame
{\it b}), we see that the crunch factor is again maximal for the
largest values of $\tan\beta$ where Yukawa couplings most nearly
unify. In frames {\it c}) and {\it d}), the higher $S$ values are
again obtained for the smallest values of $R_{tb}$ and $R_{\tau b}$, {\it i.e}
where Yukawa coupling unification most nearly occurs.

In Table 1, we show two more cases (labelled 3 and 4) 
for mass splitting only in the Higgs sector. 
In case 3, first generation scalars have masses $\sim 3000$ GeV, while
$m_{\tst_1}=589.7$ GeV and $m_{\tb_1}=581.5$ GeV. 
In this case, only the light
Higgs scalar should be accessible to Fermilab Tevatron searches, while
many sparticles 
should give observable signals at the CERN LHC $pp$
collider, operating at $\sqrt{s}=14$ TeV. In this case, $\tw_1$ and $\tz_2$
decay with unsuppressed branching fractions into three-body modes,
so that SUSY events at the LHC should be rich in
isolated leptons, as well as $b$-jets from third generation squarks
produced directly or as decay products of gluinos. 
In case 4, the first generation scalars have mass
$\sim 3300$ GeV, while the lightest third generation scalar is $\tst_1$ with 
$m_{\tst_1}=607$ GeV. The SUSY particles should again be beyond the reach
of Fermilab Tevatron experiments, but should be accessible to LHC.
The experimental signatures should again be rich in $b$-jets and 
isolated leptons produced in gluino and squark cascade decay events.

We have illustrated that the incorporation of $D$-terms allows the
construction of models with REWSB where the first two generations of
matter scalars have masses $\sim 2-3$~TeV, while other sparticle masses are 
in the sub-TeV range. 
Such a mass spectrum ameliorates (but does not completely
cure) the flavor problem associated with SUSY models. The hierarchy that
we obtain is significantly smaller that in the pioneering
papers~\cite{rimh,bfpz} where the requirement of REWSB was not
implemented. In Ref. \cite{bfpz}, the largest crunch factors are
obtained for large values ($>1$) of the unified Yukawa coupling and for
relatively small values ($10^4-10^8$~GeV) of the right handed neutrino
mass. In our study, the Yukawa coupling is typically smaller, and
(motivated by the neutrino oscillation interpretation of the atmospheric
neutrino data) we have fixed $M_N$ to be $\sim 10^{13}$~GeV. 
Various refinements to get larger values of $S$
together with REWSB are under investigation, as are the phenomenological 
consequences of IMH models.

%
\acknowledgments
This research was supported in part by the U.~S. Department of Energy
under contract number DE-FG02-97ER41022 and DE-FG-03-94ER40833.
%

%

\newpage
%
%

\iftightenlines\else\newpage\fi
\iftightenlines\global\firstfigfalse\fi
\def\dofig#1#2{\epsfxsize=#1\centerline{\epsfbox{#2}}}

\begin{table}
\begin{center}
\caption{Weak scale sparticle masses and parameters (GeV) for four IMH model
case studies. The first two cases contain full $SO(10)$ $D$-terms
applied to GUT scale SSB scalar masses. The last two cases have 
splittings applied only to the Higgs scalar masses.}
\bigskip
\begin{tabular}{lccccc}
\hline
parameter & case 1 & case 2 & case 3 & case 4 \\
\hline

$m_{16}$ & 1490.9  & 1363.9 & 2824.0 & 3239.9 \\
$m_{10}$ & 2108.5  & 1928.8 & 3993.8 & 4581.9 \\
$M_D$    & 371.7   & 276.9 & 1266.1 & 1503.7 \\
$m_{1/2}$ & 380.3  & 338.3 & 570.9 & 473.5 \\
$M_N$ & $8.03\times 10^{12}$ & $2.04\times 10^{13}$ & $3.63\times 10^{13}$ 
& $1.42\times 10^{13}$ \\
$A_0$ & -2981.8 & -2727.8  & -5648.0 & -6479.7 \\
$\tan\beta$ & 47.5 & 48.1 & 52.5 & 48.3 \\
$m_{\tg}$ & 966.9  & 869.9 & 1417.3 & 1223.3 \\
$m_{\tu_L}$ & 1716.0 & 1552.6 & 3036.3 & 3374.7 \\
$m_{\td_R}$ & 1530.8 & 1436.3 & 3022.6 & 3372.6 \\
$m_{\tell_L}$ & 1365.9 & 1296.3 & 2825.2 & 3221.4 \\
$m_{\tell_R}$ & 1553.0 & 1404.1 & 2884.4 & 3314.1 \\
$m_{\tnu_{e}}$ & 1363.6 & 1293.8 & 2824.1 & 3220.4 \\
$m_{\tst_1}$& 364.7 & 219.3 & 589.7 & 606.8 \\
$m_{\tst_2}$& 835.0 & 740.3 & 1111.3 & 1187.8 \\
$m_{\tb_1}$ & 310.9 & 405.8 & 581.5 & 918.0 \\
$m_{\tb_2}$ & 728.0 & 616.6 & 888.9 & 1027.2 \\
$m_{\ttau_1}$ & 777.8  & 515.4 & 632.2 & 949.0 \\
$m_{\ttau_2}$ & 825.3  & 732.8 & 1736.5 & 1989.8 \\
$m_{\tnu_{\tau}}$ & 780.4 & 728.0 & 1734.6 & 1988.1 \\
$m_{\tw_1}$ & 287.8 & 124.1 & 308.7 & 267.0 \\
$m_{\tz_2}$ & 288.2 & 147.2 & 316.9 & 274.9 \\
$m_{\tz_1}$ & 160.6 & 100.6 & 238.2 & 195.2 \\
$m_h$ & 125.5 & 117.6 & 111.3 & 104.3 \\
$m_A$ & 743.9 & 415.3 & 1244.0 & 2238.7 \\
$m_{H^+}$ & 750.8 & 427.0 & 1248.5 & 2242.1 \\
$\mu$ & -359.0 & 136.5 & -321.2 & 285.7 \\
$R_{tb}$ & 0.144 & 0.089 & 0.079 & 0.074 \\
$R_{\tau b}$ & 0.064 & 0.147 & 0.094 & 0.163 \\
\hline
\label{tab:cases}
\end{tabular}
\end{center}
\end{table}
%

\newpage
%

%
\begin{figure}
\dofig{5in}{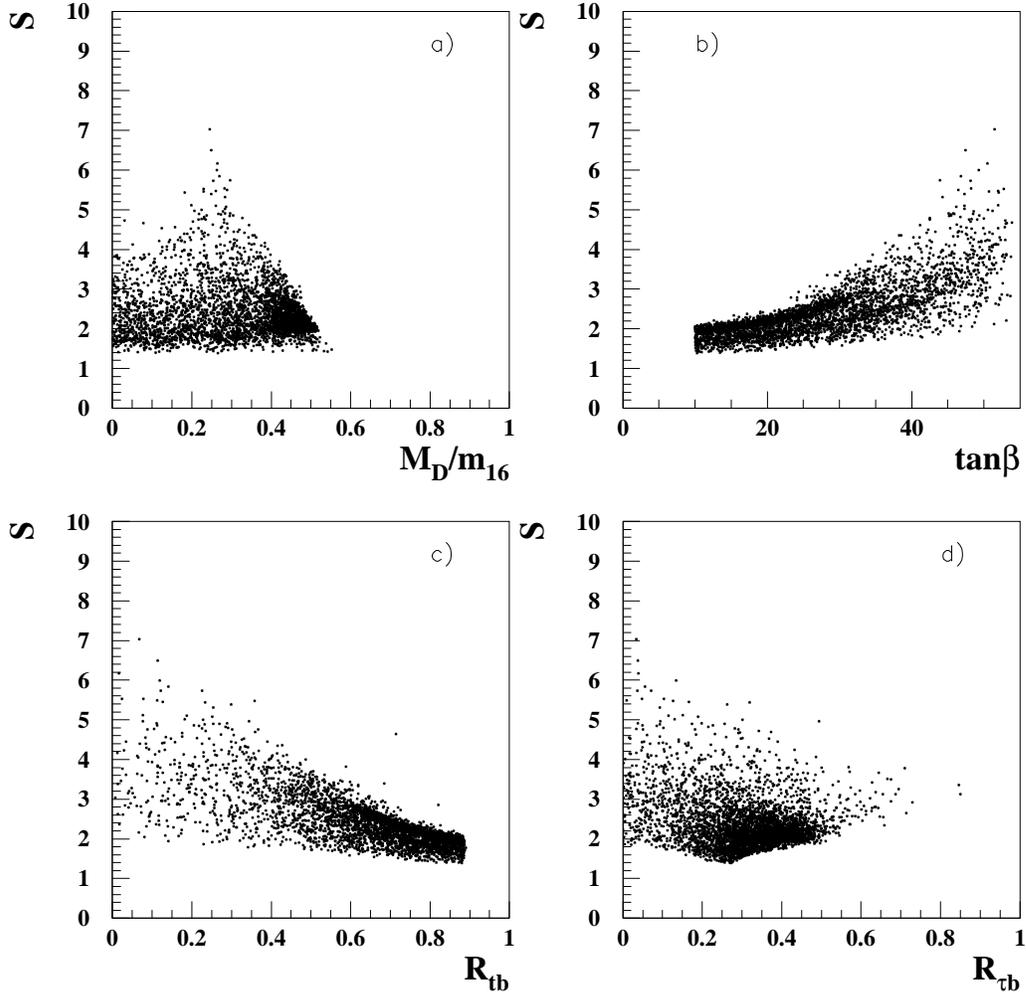}
\caption[]{ The crunch factor $S$ versus {\it a}) $M_D\over m_{16}$ and
{\it b}) $\tan\beta$. In frames {\it c}) and {\it d}), we plot the
Yukawa coupling unification ratios $R_{tb}$ and $R_{\tau b}$ defined in
the text. These models include $SO(10)$ $D$-terms, and have $\mu
<0$.}
\label{imh1}
\end{figure}
\begin{figure}
\dofig{5in}{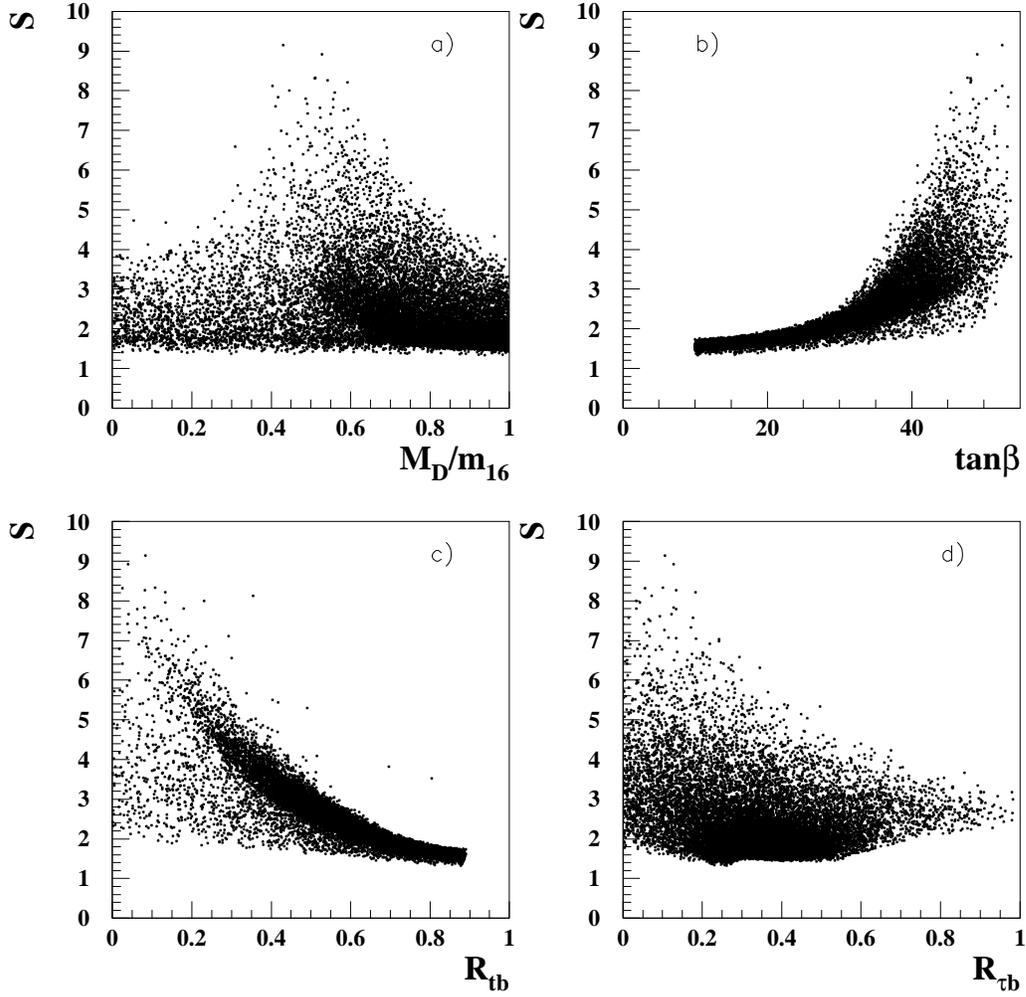}
\caption[]{
The crunch factor $S$ versus {\it a}) $M_D\over m_{16}$ and
{\it b}) $\tan\beta$. In frames {\it c}) and {\it d}), we plot the Yukawa
coupling unification ratios $R_{tb}$ and $R_{\tau b}$. These models apply
splittings only to the Higgs scalars at $M_{GUT}$. We take $\mu <0$.}
\label{imh2}
\end{figure}

\end{document}